\documentclass[letterpaper]{jpconf}
\usepackage{graphicx}
\usepackage[square,sort&compress]{natbib}
\begin{document}
\title{Antiferromagnetic behavior in CeCo$_{9}$Ge$_{4}$}

\author{C Gold$^1$, L Peyker$^1$, E-W Scheidt$^1$, H Michor$^2$ and W Scherer$^1$}

\address{$^1$ CPM, Institut f\"{u}r Physik, Universit\"{a}t Augsburg, 86159
Augsburg, Germany}
\address{$^2$ Institut f\"{u}r Festk\"{o}rperphysik, Technische Universit\"{a}t Wien,
1040 Wien, Austria}

\ead{christian.gold@physik.uni-augsburg.de}

\begin{abstract}
We investigate the novel intermetallic ternary compounds
\emph{R}Co$_{9}$Ge$_{4}$ with \emph{R} = La and Ce by means of $X$-ray diffraction, susceptibility and
specific heat measurements. CeCo$_{9}$Ge$_{4}$ crystallizes in the space group ${I}$4/${mcm}$ and is characterized by
the coexistence of two different magnetic sublattices. The Ce-based sublattice, with an
effective moment close to the expected value for a Ce$^{3+}$-ion,
exhibits a magnetically ordered ground state with $T_{\mathrm{N}}=12.5$\,K. The
Co-based sublattice, however, exhibits magnetic moments due to itinerant 3$d$ electrons.
The magnetic specific heat contribution of the Ce-sublattice is discussed in terms of
a resonance-level model implying the interplay between an antiferromagnetic phase
transition and the Kondo-effect and an underlying Schottky-anomaly indicating a
crystal field level scheme splitting into three twofold degenerated micro states ($\Delta_1 = 69$\,K, $\Delta_2 = 133$\,K).
\end{abstract}

\section{Introduction}

Cerium based intermetallic compounds of the tetragonal
Ce\emph{T}$_{9}$\emph{X}$_{4}$ (\emph{T}: transition metal, \emph{X}: Group 14 element) family
exhibit a rich variety of ground states, such as unprecedented Fermi-liquid (FL)
behavior in CeNi$_{9}$Ge$_{4}$ ($\gamma = C/T\approx5.5$\,Jmol$^{-1}$K$^{-2}$) \cite{Michor04, Killer04},
model type Kondo-lattice behavior in CeNi$_{9}$Si$_{4}$ \cite{Michor03} and valence
 fluctuations in CeCo$_{9}$Si$_{4}$ \cite{ElHagary05, Wang07}.
The most outstanding feature of CeNi$_{9}$Ge$_{4}$ is the
approximate scaling of the magnetic specific heat and susceptibility
contributions within the Ce-content in the solid solution
Ce$_{1-y}$La$_{y}$Ni$_{9}$Ge$_{4}$. This indicates that the huge
Sommerfeld coefficient $\gamma$ is mainly due to a single-ion
effect \cite{Killer04} caused by an effectively fourfold degenerate ground state in
CeNi$_{9}$Ge$_{4}$ \cite{Peyker09}. To gain insight into the spin fluctuation dynamics
and to study the influences of the crystal electrical field (CEF) of the ground
state of CeNi$_{9}$Ge$_{4}$, substitution experiments on the ligand sites are a valuable tool.\\
Control of hybridization strength, between the 4$f$ electrons
and the conduction electrons was achieved by a systematic Ni/Cu
substitution in CeNi$_{9-x}$Cu$_{x}$Ge$_{4}$. As a result a quantum critical phase
(QCP) transition at $x = 0.4$ is induced. This QCP transition is not only driven by the competition
between Kondo-effect and RKKY interaction, but also by a continuous
reduction of the effective crystal field ground state degeneracy
from non-magnetic fourfold in CeNi$_{9}$Ge$_{4}$ towards a magnetically
ordered twofold one in CeNi$_{8}$CuGe$_{4}$ \cite{Peyker09}.\\
The consequences of a Ni/Cu substitution which lead to a continuous increase of  the $d$-electron
density of state at the Fermi-level have been studied already \cite{Peyker09}.
In this paper, however, we analyzed the reverse scenario to identify the consequences of a
decreasing $d$-electron count. CeCo$_{9}$Ge$_{4}$ seems to be a suitable candidate for this study.

\section{Sample Preparation and Structural Characterization}

Polycrystalline  samples with nominal compositions of
CeCo$_{9}$Ge$_{4}$ and LaCo$_{9}$Ge$_{4}$ were prepared by a
two-step-arc-melting-procedure of the pure elements (Ce:
4N, La: 3N8, Co: 4N8, Ge: 5N) under a protective argon atmosphere.
\begin{figure}
\begin{center}
\includegraphics[width=30pc]{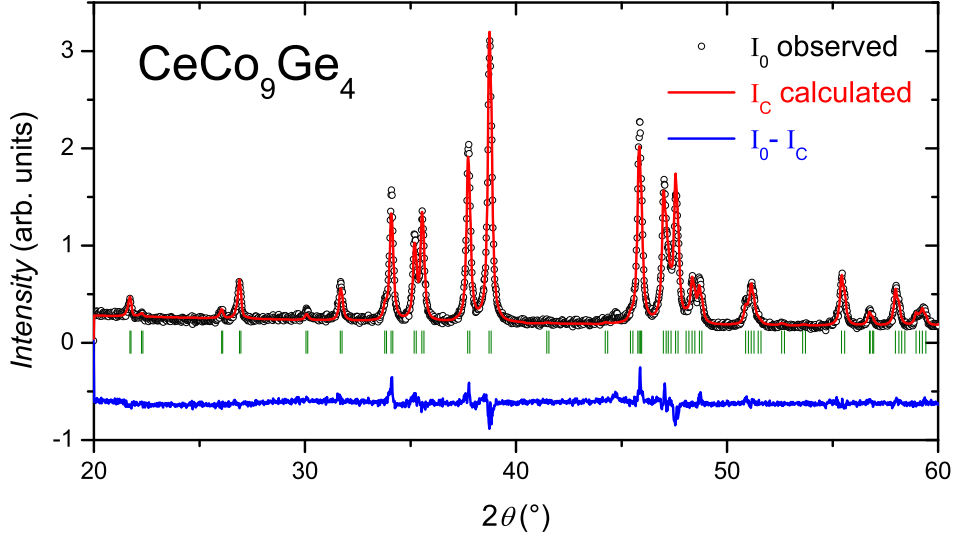}
\caption{\label{fig1} X-ray powder diffraction pattern of annealed CeCo$_{9}$Ge$_{4}$. Included is a calculated fit (Rietveld refinement) to the data, the difference
between fit and data, and tics indicating Bragg peak positions.}
\end{center}
\end{figure}
To obtain the highest possible homogeneity, the samples were turned upside-down,
remelted several times and finally annealed in evacuated quartz glass tubes for
two weeks at 950$^{\circ}$\,C. The weight losses of both samples
after annealing were less than 0.5\,$\%$ of the total mass.\\
Phase purity and crystal structure analysis were performed by means of
X-ray powder diffractometry at room temperature.
Rietveld analysis of the X-ray data confirmed that both samples crystallize in the
tetragonal LaFe$_{9}$Si$_{4}$-type \cite{Tang94} (space group ${I}$4/${mcm}$)
structure. Here, the rare-earth elements (Ce, La) occupy the
crystallographic 4$a$ sites, while the Ge atoms are situated on
the 16$l$  and the transition elements are distributed over the
16$k$, 16$l$ and 4$d$ sites. The high quality of the refinement
($R_{f}$ = 6.51) is reflected in the rather featureless difference plot in Fig.\,\ref{fig1}
for observed and calculated Bragg intensities in CeCo$_{9}$Ge$_{4}$. The resulting lattice
parameters for CeCo$_{9}$Ge$_{4}$ ($a$ = $b$ = 7.9809(1){\AA} and $c$ = 11.8825(7){\AA})
and for LaCo$_{9}$Ge$_{4}$ ($a$ = $b$ = 7.9827(8){\AA} and $c$ =
11.8743(7){\AA}) reveal a volume increase of about 8.0\% and  6.6\% in
comparison to isostructural CeCo$_{9}$Si$_{4}$
and LaCo$_{9}$Si$_{4}$ \cite{Wang07}, respectively. This pronounced
volume expansion might not only arise from differences in the
(ionic) radii of Si versus Ge. An additional volume increase
may be due to a change of the Ce-valence from mixed-valence- (CeCo$_{9}$Si$_{4}$ \cite{ElHagary05}) to a dominant Ce$^{3+}$-state in CeCo$_{9}$Ge$_{4}$.

\section{Magnetic Measurements}
The temperature dependence of the $DC$ magnetic susceptibility
$\chi(T)$ of CeCo$_{9}$Ge$_{4}$ and LaCo$_{9}$Ge$_{4}$ is shown in
Fig.\,\ref{fig2} between 2\,K and 400\,K in an external field of 0.5 T.
Above 200\,K the susceptibilities follow a modified Curie-Weiss type
law, $\chi(T) = C/(T - \Theta_\mathrm {CW}) + \chi_{0}$ (see insert Fig.\,\ref{fig2}).
From the Curie constants $C$ the effective magnetic moments
$\mu_\mathrm{eff} = 5.21 \mu_\mathrm{B}$ and
$4.54 \mu_\mathrm{B}$ were determined for the Ce- and La-compounds, respectively. For
CeCo$_{9}$Ge$_{4}$ these results lead to two different magnetic sublattices,
one is based on the local magnetic moments of the Ce$^{3+}$-ion with
$\mu_\mathrm{eff} = \sqrt{
{\mu_\mathrm{eff}^2}_{(\mathrm{CeCo}_{9}\mathrm{Ge}_{4})} -
{\mu_\mathrm{eff}^2}_{(\mathrm{LaCo}_{9}\mathrm{Ge}_{4})}}= 2.56 \mu_\mathrm{B}$, and the
second one corresponds to the Co sublattice yielding a
$\mu_\mathrm{eff}$ of $4.54 \mu_\mathrm{B}$. Assuming
the presence of Co$^{2+}$-ions displaying a 3$d^7$ configuration,
the observed effective moment for the Co sublattice implies
that only one Co-site exhibits a magnetic moment. This is in contrast
to the observation of a quasi linear magnetization curve $M(H)$, which indicates a
significant itinerant contribution. A similar behavior is observed in the
related novel 1-9-4 compound LaCo$_{9}$Si$_{4}$ \cite{Michor004},
where a strong Stoner-enhanced Pauli paramagnetism is discussed.
For LaCo$_{9}$Ge$_{4}$ the low-temperature flattening of the
susceptibility ($T < 7$\,K) corroborates the itinerant scenario.
Furthermore, the low temperature susceptibility $\chi$ for CeCo$_{9}$Ge$_{4}$ reveals
a maximum around 14\,K, indicating an antiferromagnetic phase transition
with $T_\mathrm {N}=$\,12.5\,K.
\section{Specific Heat Measurements }
\begin{figure}
\begin{minipage}{18pc}
\begin{center}
\includegraphics[width=15pc]{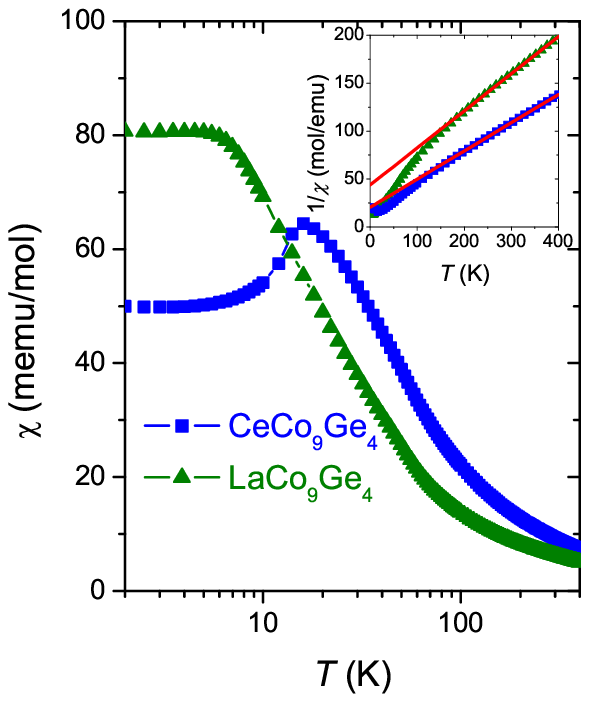}
\end{center}
\caption{\label{fig2}  The magnetic $DC$ susceptibility $\chi$ versus $T$ of CeCo$_{9}$Ge$_{4}$ and LaCo$_{9}$Ge$_{4}$ measured in an external field of 0.5\,T. Insert: Inverse magnetic $DC$ susceptibility $1/\chi$ versus $T$ of CeCo$_{9}$Ge$_{4}$ (squares) and LaCo$_{9}$Ge$_{4}$ (triangles). The solid lines are fits based on a modified Curie-Weiss type law, $\chi(T) = C/(T - \Theta_\mathrm {CW}) + \chi_{0}$.}
\end{minipage}\hspace{2pc}%
\begin{minipage}{18pc}
\begin{center}
\includegraphics[width=18pc]{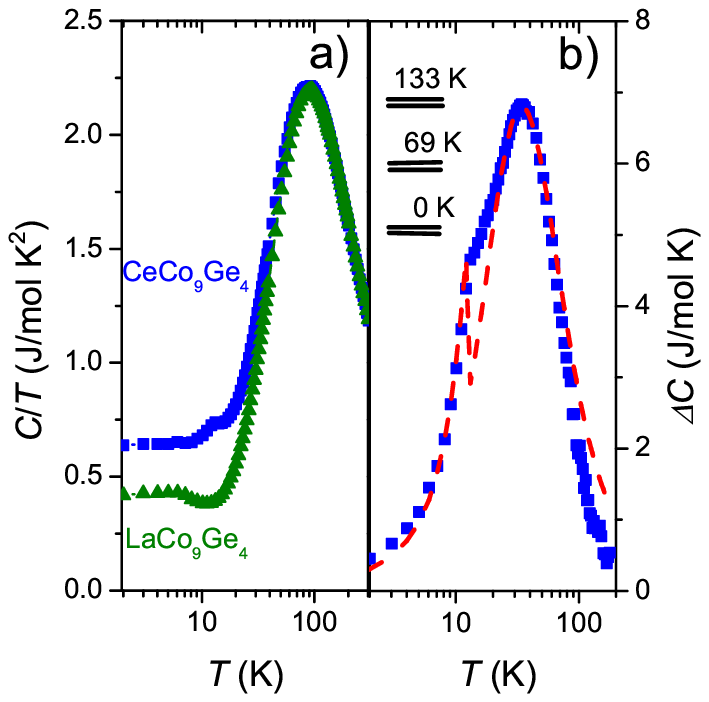}
\end{center}
\caption{\label{fig3} a) The temperature dependent specific heat divided by the temperature $C/T$ of CeCo$_{9}$Ge$_{4}$ and LaCo$_{9}$Ge$_{4}$.
b) The magnetic contribution of the specific heat $\Delta$$C$ of CeCo$_{9}$Ge$_{4}$. The dashed line represents a theoretical adjustment
to the data taking into account the resonant-level model \cite{Schotte75} and a Schottky anomaly.}
\end{minipage}
\end{figure}
The temperature dependent specific heat $C(T)$ divided by
temperature of CeCo$_{9}$Ge$_{4}$ and LaCo$_{9}$Ge$_{4}$ between 2\,K
to 300\,K is displayed in Fig.\,\ref{fig3}a. Besides the high temperature regime
of the specific heat, which is dominated by phonon contributions $C(T)/T$ of
CeCo$_{9}$Ge$_{4}$ exhibits also a pronounced anomaly at 12.5 K
indicating a phase transition to an ordered magnetic state, in
good agreement with the $\chi(T)$ data. By contrast $C(T)$ of
LaCo$_{9}$Ge$_{4}$ does not feature any evidence of magnetic
order but a small increase below 10\,K in correspondence to the
observed flattening in $\chi(T)$.\\
Due to the fact, that itinerant 3$d$ magnetism is presented in
both samples, it is difficult to quantify the linear
Sommerfeld coefficient $\gamma$. In order to disregard
this effect and to eliminate the phonon contribution, we subtracted
the total specific heat of the isostructural LaCo$_{9}$Ge$_{4}$ sample
from the Ce compound. As a result we obtained the magnetic contribution
of the specific  heat $\Delta$$C$ of the Ce subsystem pictured in Fig.\,\ref{fig3}b.
From a $\Delta$$C /T$ over $T^2$ plot a Sommerfeld coefficient
$\gamma\approx 200$\,mJmol$^{-1}$K$^{-2}$ is extracted, yielding a Kondo-temperature
of $T_{\mathrm K} =0.68 R/ \gamma = 28$\,K, where $R$ is the gas constant.
This pronounced Kondo-contribution at low temperatures can
be described utilizing the resonant-level model \cite{Schotte75}
in combination with a molecular field
approach to account for long-range magnetic order
(see, e.g., \cite{Peyker09}). The results of these calculations
qualitatively reproduce the evolution of the magnetic specific heat
anomaly ($T_{\mathrm N}$ = 12.5 K) and the Kondo-contribution of
CeCo$_{9}$Ge$_{4}$ (see dashed line in Fig.\,\ref{fig2}b). The exchange
interaction $J$ and the Kondo-temperature $T_{\mathrm K}$, obtained
from the resonant-level model, are 65\,K and 32\,K in good agreement
with the $T_{\mathrm K}$-value obtained from the  $\gamma$-value.
In addition, at high temperatures the specific heat data
can be well described by a Schottky anomaly resulting  from a CEF  effect
of the $J = 5/2$ state of the Ce$^{3+}$-ion. The
Schottky maximum can be fitted using a crystal-field level scheme with
energies levels separations of $\Delta_1 = 69$\,K and $\Delta_2 = 133$\,K.
The entropy calculation from $\Delta C/T$
nearly tends above 130\,K towards the expected value of $R$ln6,
characteristic for a sixfold degenerated system.

\section{Conclusion }
The influence of the observed volume expansion of about 1.2\% as well as the reduction
of the $d$-electron count compared to CeNi$_{9}$Ge$_{4}$ was investigated using susceptibility
and specific heat measurements. We have found a
coexistence of two magnetic sublattices in CeCo$_{9}$Ge$_{4}$. While
the Ce-sublattice is characterized by an antiferromagnetic ordered Kondo-lattice
($T_{\mathrm N}$ = 12.5\,K) with a Sommerfeld value of $\gamma \simeq 200
 $mJ/mol K$^{2}$ and a Kondo-temperature of $T_{\mathrm K}$ = 32 K,
the Co-sublattice reveals 3$d$ itinerant paramagnetism.
In addition, the Ni/Co exchange leads to a shift of the CEF levels. The
isostructural compound LaCo$_{9}$Ge$_{4}$ is, due to the missing
magnetic rare-earth sublattice, an ideal system to study the role of
itinerant magnetism in CeCo$_{9}$Ge$_{4}$ and will be subject of future investigations.

\section{Acknowledgments}
We acknowledge valuable discussions with E Bauer. This work was supported by the Deutsche Forschungsgemeinschaft (DFG)
under Contract No. SCHE487/7-1 and by the COST P16 ECOM project of
the European Union.

\providecommand{\newblock}{}

\end{document}